\documentclass[fleqn,usenatbib]{mnras}

\usepackage{graphicx}
\usepackage{ifthen}
\usepackage{url}
\usepackage{amsmath}
\usepackage{bm}
\usepackage{lscape}
\usepackage{rotating}
\usepackage[graphicx]{realboxes}
\usepackage{supertabular}
\usepackage{pdfpages}
\usepackage{footnote}
\usepackage{hyperref}
\usepackage{amssymb}
\makesavenoteenv{tabular}
\makesavenoteenv{table}

\usepackage[T1]{fontenc}
\usepackage{tipa}
\usepackage{accents}
\newcommand{\ubar}[1]{\underaccent{\bar}{#1}}
\usepackage{tablefootnote}

\hypersetup{
  colorlinks   = true,    
  urlcolor     = blue,    
  linkcolor    = blue,    
  citecolor    = blue      
}

\usepackage{ulem}

\def\ltsima{$\; \buildrel < \over \sim \;$}
\def\lta{\lower.5ex\hbox{\ltsima}}
\def\gtsima{$\; \buildrel > \over \sim \;$}
\def\simgt{\lower.5ex\hbox{\gtsima}}
\def\kms{{\rm\,km \  s^{-1}}}

\def\kpc{{\rm\,kpc}}

\def\msun{{\rm\,M_\odot}}

\usepackage{amsmath}

\def\s{\ifmmode \widetilde \else \~\fi}
\def\={\overline}

\def\spose#1{\hbox to 0pt{#1\hss}}

\def\eg{{e.g.,\ }}
\def\ie{{i.e.,\ }}
\def\lta{\mathrel{\spose{\lower 3pt\hbox{$\mathchar"218$}}
     \raise 2.0pt\hbox{$\mathchar"13C$}}}
\def\gta{\mathrel{\spose{\lower 3pt\hbox{$\mathchar"218$}}
     \raise 2.0pt\hbox{$\mathchar"13E$}}}
\def\Dt{\spose{\raise 1.5ex\hbox{\hskip3pt$\mathchar"201$}}}    
\def\dt{\spose{\raise 1.0ex\hbox{\hskip2pt$\mathchar"201$}}}    

\def\dotsfill{\leaders\hbox to 1em{\hss.\hss}\hfill}

\def\FeH{{\rm[Fe/H]}}

\def\ione{{~\sc i}}
\def\ii{{~\sc ii}}

\loadboldmathitalic 
\title[The GHOST tale of a PIGS star]{GHOST Commissioning Science Results II:  a very metal-poor star witnessing the early Galactic assembly} 
\author[F. Sestito et al.] {Federico Sestito$^{1}$\thanks{Email: \url{sestitof@uvic.ca}}, 
Christian R. Hayes$^{2}$,
Kim A. Venn$^{1}$,
Jaclyn Jensen$^{1}$,
\newauthor
Alan W. McConnachie$^{2,1}$, 
John Pazder$^{2,1}$,
Fletcher Waller$^{1}$,
Anke Ardern-Arentsen$^{3}$, 
\newauthor
Pascale Jablonka$^{4,5}$,
Nicolas F. Martin$^{6,7}$,
Tadafumi Matsuno$^{8}$,
Julio F. Navarro$^{1}$,
\newauthor
Else Starkenburg$^{8}$,
Sara Vitali$^{9}$,
John Bassett$^{10}$,
Trystyn A. M. Berg$^{11,2}$,
Ruben Diaz$^{10}$,
\newauthor
Michael L. Edgar$^{12}$,
Veronica Firpo$^{10}$,
Manuel Gomez-Jimenez$^{10}$,
Venu Kalari$^{10}$,
\newauthor
Sam Lambert$^{2}$,
Jon Lawrence$^{13}$,
Gordon Robertson$^{13}$,
Roque Ruiz-Carmona$^{10}$,
\newauthor
Ricardo Salinas$^{10}$,
Kim M. Sebo$^{14}$,
and 
Sudharshan Venkatesan$^{13}$
\\
$^{1}$ Department of Physics and Astronomy, University of Victoria, PO Box 3055, STN CSC, Victoria BC V8W 3P6, Canada\\
$^{2}$ NRC Herzberg Astronomy \& Astrophysics, 5071 West Saanich Road, Victoria, BC V9E 2E7, Canada\\
$^{3}$ Institute of Astronomy, University of Cambridge, Madingley Road, Cambridge CB3 0HA, UK \\
$^{4}$ Laboratoire d'astrophysique, \'Ecole Polytechnique F\'ed\'erale de Lausanne (EPFL), Observatoire, CH-1290 Versoix, Switzerland\\
$^{5}$ GEPI, Observatoire de Paris, Universit\'e PSL, CNRS, 5 Place Jules Janssen, F-92195 Meudon, France\\
$^{6}$ Universit\'e de Strasbourg, CNRS, Observatoire astronomique de Strasbourg, UMR 7550, F-67000 Strasbourg, France\\
$^{7}$ Max-Planck-Institut f\"{u}r Astronomie, K\"{o}nigstuhl 17, D-69117 Heidelberg, Germany \\
$^{8}$ Kapteyn Astronomical Institute, University of Groningen, Landleven 12, NL-9747AD Groningen, the Netherlands\\
$^{9}$ Instituto de Estudios Astrof\'isicos, Universidad Diego Portales, Av. Ej\'ercito Libertador 441, Santiago, Chile\\
$^{10}$ Gemini Observatory/NSF’s NOIRLab, Casilla 603, La Serena, Chile\\
$^{11}$ Dipartimento di Fisica G. Occhialini, Universit\'a degli Studi di Milano Bicocca, Piazza della Scienza 3, I-20126 Milano, Italy\\
$^{12}$ Australian Astronomical Observatory\\
$^{13}$ Australian Astronomical Optics, Macquarie University, 105 Delhi Rd, North Ryde NSW 2113, Australia\\
$^{14}$ Research School of Astronomy and Astrophysics, College of Science, Australian National University, Canberra 2611, Australia
}

\pubyear{2024}
\date{Accepted XXX. Received YYY; in original form ZZZ}
\begin{document}
\maketitle 
\label{firstpage}
\pagerange{\pageref{firstpage}--\pageref{lastpage}}

\begin{abstract}
This study focuses on Pristine$\_180956.78$$-$$294759.8$ (hereafter P180956, $\FeH=-1.95\pm0.02$), a star selected from the Pristine Inner Galaxy Survey (PIGS), and followed-up with the recently commissioned Gemini High-resolution Optical SpecTrograph (GHOST) at the Gemini South telescope. The GHOST spectrograph's high efficiency in the blue spectral region ($3700-4800$~\AA) enables the detection of elemental tracers of early supernovae  (\eg Al, Mn, Sr, Eu). The star exhibits chemical signatures resembling those found in ultra-faint dwarf systems, characterised by very low abundances of neutron-capture elements (Sr, Ba, Eu), which are uncommon among stars in the Milky Way halo. Our analysis suggests that  P180956 bears the chemical imprints of a small number (2 or 4) of low-mass hypernovae ($\sim10-15\msun$), which are needed to mostly reproduce the abundance pattern of the light-elements (\eg [Si, Ti/Mg, Ca] $\sim0.6$), and one fast-rotating intermediate-mass supernova ($\sim300\kms$, $\sim80-120\msun$), which is the main channel contributing to the high [Sr/Ba] ($\sim +1.2$). The small pericentric ($\sim0.7\kpc$) and apocentric ($\sim13\kpc$) distances and its orbit confined to the plane ($\lesssim 2\kpc$), indicate that this star was likely accreted during the early Galactic assembly phase. Its chemo-dynamical properties  suggest that  P180956 formed in a system similar to an ultra-faint dwarf galaxy accreted either alone, as one of the low-mass building blocks of the proto-Galaxy, or as a satellite of Gaia-Sausage-Enceladus. The combination of Gemini's large aperture with GHOST's high efficiency and broad spectral coverage makes this new spectrograph one of the leading instruments for near-field cosmology investigations.
\end{abstract}


\section{Introduction}
Low-metallicity stars are among the oldest stars in the Galaxy. Cosmological simulations suggest that  most metal-poor stars formed within $2-3$ Gyr after the Big Bang, likely in low-mass systems that were accreted early on into the Galaxy \citep["building blocks", \eg][]{Starkenburg17a, ElBadry18,Sestito21}. These merging building blocks contributed stars, gas, and dark matter to the proto-Milky Way. Because they formed at the bottom of the potential well of the merging building blocks, these stars are predicted to occupy the inner regions of the present-day Galaxy \citep[\eg][]{White00,Starkenburg17a, ElBadry18, Sestito21}. Systems  accreted later are anticipated to disperse their stars primarily in the halo \citep{Bullock2005, Johnston2008, Tissera12}, or possibly the disc \citep[\eg][]{Abadi03, Sestito21, Santistevan21}. An in-situ component may also form from the gas deposited by the building blocks, recently called the Aurora stars \citep{Belokurov22}.  This in-situ component may have formed in a lumpy and chaotic interstellar medium (ISM), possibly resembling the chemical properties of globular clusters \citep{Belokurov23}.

The most chemically pristine stars in the Milky Way (MW) may have been enriched by only one or a few  supernovae or hypernovae events \citep[\eg][]{Frebel10,Ishigaki18}. The study of the orbital properties and chemical  abundance patterns of pristine stars is essential for understanding the lives and deaths of the first stars and the assembly history of the Galaxy  \citep[\eg][]{Freeman02, Venn04, Tumlinson10, Wise12, Karlsson13}. 

Metal-poor stars in and towards the Galactic bulge can serve as important tracers of the earliest stages of Galactic assembly, yet their detection is extremely challenging \citep[e.g.,][]{Schlaufman14, Lamb2017}. The inner regions of the MW are dominated by a metal-rich population and disrupted globular clusters \citep{Ness13a, Ness2014, Bensby13, Bensby17, Schiavon17, Schultheis19}. Furthermore, extreme interstellar extinction and stellar crowding have made photometric surveys of bulge metal-poor stars exceedingly difficult.

The Abundances and Radial velocity Galactic Origins Survey  \citep[ARGOS,][]{Ness13b} found that $\lesssim1$ percent of their sample had \FeH\footnote{\FeH $= \log(\rm{N_{Fe}/N_{H}}){\star}-\log(\rm{N_{Fe}/N_{H}})_{\odot} $, where $\rm{N_X}$ represents the number density of element X.}~$<-1.5$, resulting in a total of 84 stars. The metallicity-sensitive photometric filter from the SkyMapper Southern Survey \citep{Bessell11,Wolf18} has been used by the  Extremely Metal-poor BuLge stars with AAOmega  \citep[EMBLA,][]{Howes14,Howes15,Howes16} survey to select very metal-poor stars (VMPs, $\FeH\leq-2.0$). Their high-resolution analysis of 63 VMPs revealed that the majority resembled chemically metal-poor stars in the Galactic halo, with the exception of a lack of carbon-rich stars and a larger scatter in [$\alpha$/Fe] abundances. Additionally, their kinematic analysis found that it was challenging to distinguish stars that were born in the bulge from those that are merely in the inner halo.
The Chemical Origins of Metal-poor Bulge Stars \citep[COMBS,][]{Lucey19} studied the chemo-dynamical properties of  inner Galactic stars, finding that around $\sim50$ percent of their sample is composed of halo interlopers, while their chemical properties resemble those of the halo \citep{Lucey21,Lucey22}.

Similar to the EMBLA survey, the Pristine Inner Galaxy Survey \citep[PIGS,][]{Arentsen20a,Arentsen20b} selected metal-poor targets from the narrow-band photometry of the  Pristine survey \citep{Starkenburg17b}. The Pristine survey, conducted at the Canada-France-Hawaii Telescope (CFHT), utilises the CaHK filter in combination with broad-band photometry to provide a highly efficient method of identifying low-metallicity stars \citep[$\sim$56 percent success rate at $\FeH\leq-2.5$,][]{Youakim17, Aguado19, Venn20,Lucchesi22}. Around $\sim$12,000 inner Galaxy metal-poor  candidates  selected by PIGS  were observed with low-/medium-resolution spectroscopy using the AAOmega spectrograph on the Anglo Australian Telescope (AAT). The results of these observations  showed   $\sim$80 percent efficiency in identifying VMP stars towards the bulge \citep{Arentsen20b} and the Sagittarius dwarf galaxy \citep{Vitali22} using the Pristine metallicity-sensitive filter for initial selection. Interestingly, within PIGS, \citet{Mashonkina23} reports the serendipitous discovery of the first  r- and s- processes rich Carbon-enhanced star (CEMP$-$r/s) in the inner Galaxy.

In a recent study, \citet{Sestito23} analysed high-resolution spectra of 17 metal-poor stars selected from the PIGS survey taken with the Gemini Remote Access to CFHT ESPaDOnS Spectrograph \citep[GRACES,][]{Chene14,Pazder14}. Their findings, consistent with \citet{Howes16}, indicate that the chemo-dynamical properties of the VMP population in the inner Galaxy resemble that of the halo, suggesting a common origin from disrupted building blocks. \citet{Sestito23} report stars with chemical abundances compatible with those of disrupted second-generation globular clusters (GC), one with exceptionally low metallicity ($\FeH\sim-3.3$), well below the metallicity floor of GCs \citep[$\FeH\sim-2.8$,][]{Beasley19}. This provides further evidence that extremely metal-poor structures (EMPs, $\FeH\leq-3.0$) can form in the early Universe \citep[see also][on the discovery of the disrupted EMP globular cluster, C-19]{Martin22}.

In this paper, we present a new analysis of the inner Galactic very metal-poor Pristine$\_180956.78$$-$$294759.8$ (P180956), from spectra taken during the commissioning of the new Gemini High-resolution Optical SpecTrograph \citep[GHOST,][]{Pazder20}.  This star was previously analysed in the PIGS/GRACES analysis \citep{Sestito23},  but it can be observed from either the northern hemisphere (GRACES) or the south (GHOST, at Gemini South). P180956 is selected for this work for its unusually low [Na, Ca/Mg] and [Ba/Fe] ratios and for its high eccentric orbit that remains confined close to the Milky Way plane. The new GHOST spectrograph has also been used in the analysis of two stars in the Reticulum~II ultra-faint dwarf galaxy \citep{Hayes23}, and the presentation of the spectrum of the r-process rich standard star HD\,222925 \citep{Hayes22, McConnachie22}. GHOST's  wide spectral coverage and its  high efficiency in the blue region are crucial to detect species that are tracers of the early chemical evolution (\eg Al, Mn, Sr, Eu) and that were not accessible with GRACES spectrograph. 

The instrument setup and data reduction are described in Section~\ref{sec:data}. Sections~\ref{sec:models}~and~\ref{sec:chems} discuss the model atmospheres analysis and the chemical abundance analyses, respectively. Orbital parameters are reported in Section~\ref{sec:orbit}. The results are discussed in Section~\ref{sec:discussion}, focusing on the type of supernovae that polluted the formation site of P180956, and its origin in an ancient dwarf system, which may have resembled today's ultra-faint dwarfs.  Conclusions are presented in Section~\ref{sec:conclusions}.

\section{Data}\label{sec:data}

\subsection{GHOST observations} 
The target (G $=13.50$ mag), P180956, was initially observed as part of the PIGS photometric survey using MegaCam at the Canada-France-Hawaii Telescope (CFHT).  Its small pericentric distance (r$_{\rm{peri}} \sim 0.7 \kpc$), the apocentre ($\sim13\kpc$), the limited maximum excursion from the plane (Z$_{\rm{max}} \sim 1.8 \kpc$), and its high eccentricity ($\epsilon\sim 0.90$) imply that it was likely accreted during the early stages of Galactic assembly \citep{Sestito23}. A low/medium-resolution (R $\sim1,300$ and R $\sim11,000$) spectrum of P180956 was obtained using AAT/AAOmega   \citep{Arentsen20b}, and also with the Gemini Remote Access to CFHT ESPaDOnS Spectrograph \citep[GRACES, R $\sim40,000$,][]{Chene14,Pazder14} as part of LLP-102 (PI K.A. Venn). The GRACES spectrum was analysed by \citet{Sestito23} along with 16 other VMP stars in the Galactic bulge; however, due to limitations in the throughput of the 300-metre optical fiber, the bluest spectral regions were not accessible for that study.  The GRACES spectral analysis was limited to a range of $4900-10000$ \AA{} only. Thus, P180956 was chosen as a commissioning target for the Gemini High-resolution Optical SpecTrograph \citep[GHOST,][]{Pazder20}. The instrument's high efficiency in the blue spectral region makes it ideal for detecting spectral features of additional chemical elements which allow a deeper investigation of the origins of P180956.

The target was observed on September 12th, 2022, during the second commissioning run of GHOST. Three exposures, each lasting 600 seconds, were conducted. The instrument was configured in the standard resolution (R $\sim50,000$) single object mode, employing a spectral and spatial binning of 2 and 4, respectively. The instrument setup allows to cover the $3630-5440$ \AA{} region with the blue arm and the $5210-9500$ \AA{} region with the red arm. This specific setup and exposure times were chosen to enable the detection of spectral lines from species in the bluer regions ($\sim$4000 \AA) of the spectrum that were not accessible with GRACES, \ie C, Al, Si, Sc, V, Mn, Co, Cu, Zn, Sr, Y, La, Eu. Table~\ref{tab:obs} reports the {\it Gaia} DR3 source ID and  photometry, the reddening from \citet{Green19} along with the total exposure time and the number of exposures.

\begin{table}
\caption{Log of the observations. The Pristine name, the short name, the  source ID,  G, and BP$-$RP from  {\it Gaia} DR3, the reddening from the 3D map of \citet{Green19}, the total exposure time, the number of exposures, the SNR, and the radial velocity are reported. The  SNR is measured as the ratio between the median flux and its standard deviation in three spectral regions, close to the Eu\ii{} 4129 \AA{}, in the Mg\ione b Triplet and in the Na\ione{} Doublet.}
\label{tab:obs}
\resizebox{0.5\textwidth}{!}{
\begin{tabular}{lr}
\hline
Property & Value\\
\hline
Pristine name & Pristine$\_180956.78$$-$$294759.8$ \\
source ID   & 4050071013878221696\\
ra & 272.4865677802622\\
dec & -29.799952273156627\\
G & 13.50 mag\\
BP$-$RP & 1.55  mag\\
E(B$-$V) & 0.53 mag\\
t$_{\rm exp}$& 1800 s\\
N$_{\rm exp}$ & 3\\
SNR @Eu\ii{}, @Mg\ione b, @Na\ione{}   & 16, 50, 65\\
RV & $262.976 \pm 0.363$ $\kms$\\
T$_{\rm{eff}}$ & $5391 \pm 133$ K\\
logg & $1.87 \pm 0.10$\\
\hline
\hline
\end{tabular}
}
\end{table}

\subsection{Data reduction}
The acquired spectra were processed using the GHOST Data Reduction pipeline \citep[\textsc{GHOSTDR},][]{Ireland18,Hayes22}, which is integrated into the \textsc{DRAGONS} suite \citep{Labrie19}. This pipeline\footnote{\url{https://github.com/GeminiDRSoftware/GHOSTDR}} generates 1D spectra for the blue and red arms, which were wavelength calibrated, order-combined, and sky-subtracted. Subsequently, the spectra were individually normalized through polynomial fitting, and barycentric corrections were applied. Radial velocities and their uncertainties have been measured for the blue and red arms using the \textsc{Doppler} code\footnote{\url{https://github.com/dnidever/doppler}}, which cross-correlate the observed spectra with a VMP template synthesized with \textsc{Turbospectrum}\footnote{\url{https://github.com/bertrandplez/Turbospectrum2019}} \citep{Plez12} and a \textsc{MARCS}\footnote{\url{https://marcs.astro.uu.se}} model atmosphere \citep{Gustafsson08}.
The  radial velocity measured from the blue and red arms are RV~$= 263.034 \pm 0.020 \kms$ and RV~$=262.918 \pm 0.362 \kms$, respectively. Their average, RV~$=262.976 \pm 0.363\kms$, is in agreement within $0.8\sigma$ from the measurement in the GRACES analysis \citep[RV~$=262.45\pm0.54\kms$,][]{Sestito23} and by $2.3\sigma$ with the value from the low-/medium-resolution PIGS campaign \citep[RV~$=264.86\pm0.73\kms$,][]{Arentsen20a}. The lack of RV variation likely rules out the possibility that the object is in a binary system. The continuum points in the observations were identified using spectral templates and fitted via an iterative sigma clipping method to obtain a normalised spectrum.  Finally, the blue and red output spectra were merged together with inverse-variance weighting in the overlapping regions.

\begin{figure*}
\includegraphics[width=1\textwidth]{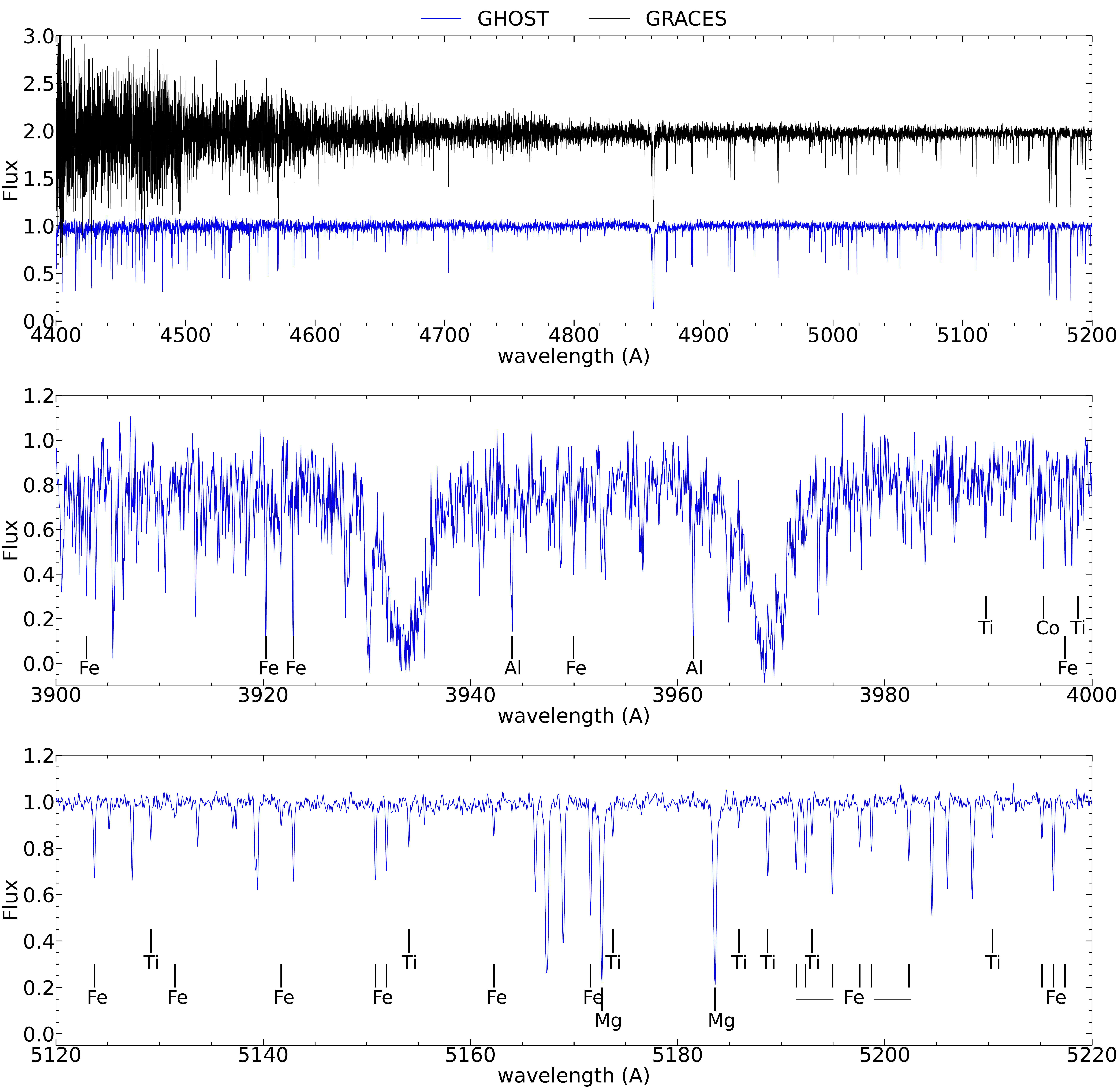}
\caption{GHOST spectrum of P180956. Top panel: Wide spectral region, $4400-5200$ \AA{}, also showing the GRACES spectrum (black line). Central panel: blue region $3900-4000$ \AA{}. Bottom panel: Mg\ione b Triplet. Spectral lines used for the chemical abundance analysis are marked.
}
\label{Fig:spectra}
\end{figure*}

Figure~\ref{Fig:spectra} showcases the reduced spectrum of P180956. In the top panel, a comparison is made with the GRACES observation in the $4400-5200$ \AA{} region. Both instruments  have similar spectral (high) resolution; however, the SNR of the GRACES spectrum (black line) deteriorates below $5000$ \AA{}, a region where the number of spectral lines clearly increases, as seen in the GHOST spectrum (blue). The Balmer line H$-\epsilon$, the Ca\ii{} H\&K lines, along with two Al\ione{} lines, two Ti\ione{}, five Fe\ione{} lines, and one Co\ione{} line are shown in the central panel. The Mg\ione b Triplet ($\sim5180$ \AA) region also contains several Fe and Ti lines, as shown in the bottom panel.  Table~\ref{tab:obs} reports  the SNR measured close to the Eu\ii{} 4129 \AA{} line, the Mg\ione b Triplet region, and to the Na\ione{} Doublet ($\sim5890$ \AA).

\section{Model Atmospheres Analysis} \label{sec:models}

\subsection{Stellar parameters}\label{sec:stellparam}

The stellar parameters used in this study are adopted from the GRACES analysis by \citet[][T$_{\rm{eff} } = 5391 \pm 133$ K, logg $=1.87\pm0.10$]{Sestito23}. To briefly summarize, the effective temperature is estimated based on the color-temperature relationship derived by \citet[][]{Mucciarelli21}. This relationship is based on the Infrared Flux Method introduced by \citet{Gonzalez09} and adapted to the {\it Gaia} EDR3 photometry. The surface gravity is determined by applying the Stefan-Boltzmann equation, inferred assuming a flat mass distribution between 0.5 to 0.8 $\msun$. These calculations rely on several factors: 1) the de-reddened photometry\footnote{Extinction is from the 3D map of \citet{Green19}. To convert from the  E(B-V) map to  {\it Gaia} extinction coefficients,  the  $\rm A_V/E(B-V)= 3.1$ \citep{Schultz75} and the $\rm A_G/A_V = 0.85926$, $\rm A_{BP} /A_V = 1.06794$, $\rm A_{RP} /A_V = 0.65199$ relations \citep{Marigo08,Evans18} are used.}; 2) the distance to the star, which is estimated to be $3.30\pm0.27$ kpc \citep{Sestito23}; and 3) a metallicity \citep[taken as $\FeH =-2.0\pm0.1$,][]{Sestito23}. Uncertainties on the stellar parameters are derived using a Monte Carlo simulation.

The microturbulence velocity (v$_{\rm{micro}}$) is obtained spectroscopically achieving a flatter distribution in the abundances from the Fe\ione{} lines, A(Fe\ione)\footnote{A(X)~$= 12 + \rm{Log}( N_X/N_H)$} vs. the reduced equivalent width. This gives a v$_{\rm{micro}} =1.5\pm0.1\kms$.

\subsection{Spectral lines and atomic data}
The spectral line list is generated with \textsc{linemake} \citep{Placco21}, including lines with hyper-fine structure corrections (Sc, V, Mn, Co, and Cu), molecular bands (CH in the 4300 \AA{} region) and r-process isotopic corrections (Ba, Eu). For CH, a ratio of $^{12}$C/$^{13}$C~$= 5$ has been assumed as for a typical RGB star \citep{Spite06}.  Solar abundances are taken from \citet{Asplund09}.

An initial measurement of the equivalent width is performed using \textsc{DAOSPEC} \citep{Stetson08}, which automatically fits Gaussian profiles to spectra following the input line list.  Given the signal-to-noise ratio of our spectrum,  lines weaker than 15 m\AA{} are rejected, and lines stronger than 100 m\AA{} are further examined with non-Gaussian measurements, \ie with an integral. The equivalent widths are then used with the \textsc{MOOG}\footnote{\url{https://www.as.utexas.edu/~chris/moog.html}} spectrum synthesis code \citep{Sneden73,Sobeck11} to determine the chemical abundances assuming Local Thermodynamic Equilibrium (LTE). The spherical \textsc{MARCS}\footnote{\url{https://marcs.astro.uu.se}} model atmospheres \citep{Gustafsson08,Plez12}, which assume [$\alpha$/Fe]~$=0.4$, are used for the chemical abundance analysis in this paper, which yield to  $\FeH = -1.95\pm0.02$.  The chemical abundances of Sc, Cu, Y, Ba, La, and Eu are determined using the \textsc{synth} mode within \textsc{MOOG}. 

The  list of spectral lines used for the chemical abundance analysis, their atomic data, their EWs and abundances are reported as supplementary online material.

\subsection{Uncertainties on the chemical abundances}
\textsc{MOOG} provides estimates of the chemical abundances A(X) along with their line-to-line scatter, $\delta_{\rm A(X)}$. The total abundance scatters, $\delta_{\rm A(X),TOT}$, are calculated by combining in quadrature the line-to-line scatter with the uncertainties resulting from variations in the stellar parameters ($\delta_{\rm T_{eff}}$, $\delta_{\rm logg}$, both $\sim0.02$), in the microturbulence velocity ($\delta_{\rm v_{micro}}\sim0.02$) and in the metallicity ($\delta_{\FeH}\sim0.02$). The final uncertainty for element X is given by $\sigma_{\rm A(X)}=\delta_{\rm A(X),TOT}/\sqrt{{\rm N_X}}$. In case there is only one spectral line or \textsc{synth} mode is employed, the dispersion in Fe\ione{} lines is considered as the typical dispersion.

\section{Chemical abundance analysis}\label{sec:chems}
The blue wavelength coverage of GHOST permits an analysis of the spectral lines of many elements in metal-poor stars, including a range of $\alpha$-, odd-Z, Fe-peak, and neutron-capture process elements. 

\subsection{Carbon}
\begin{figure}
\includegraphics[width=0.5\textwidth]{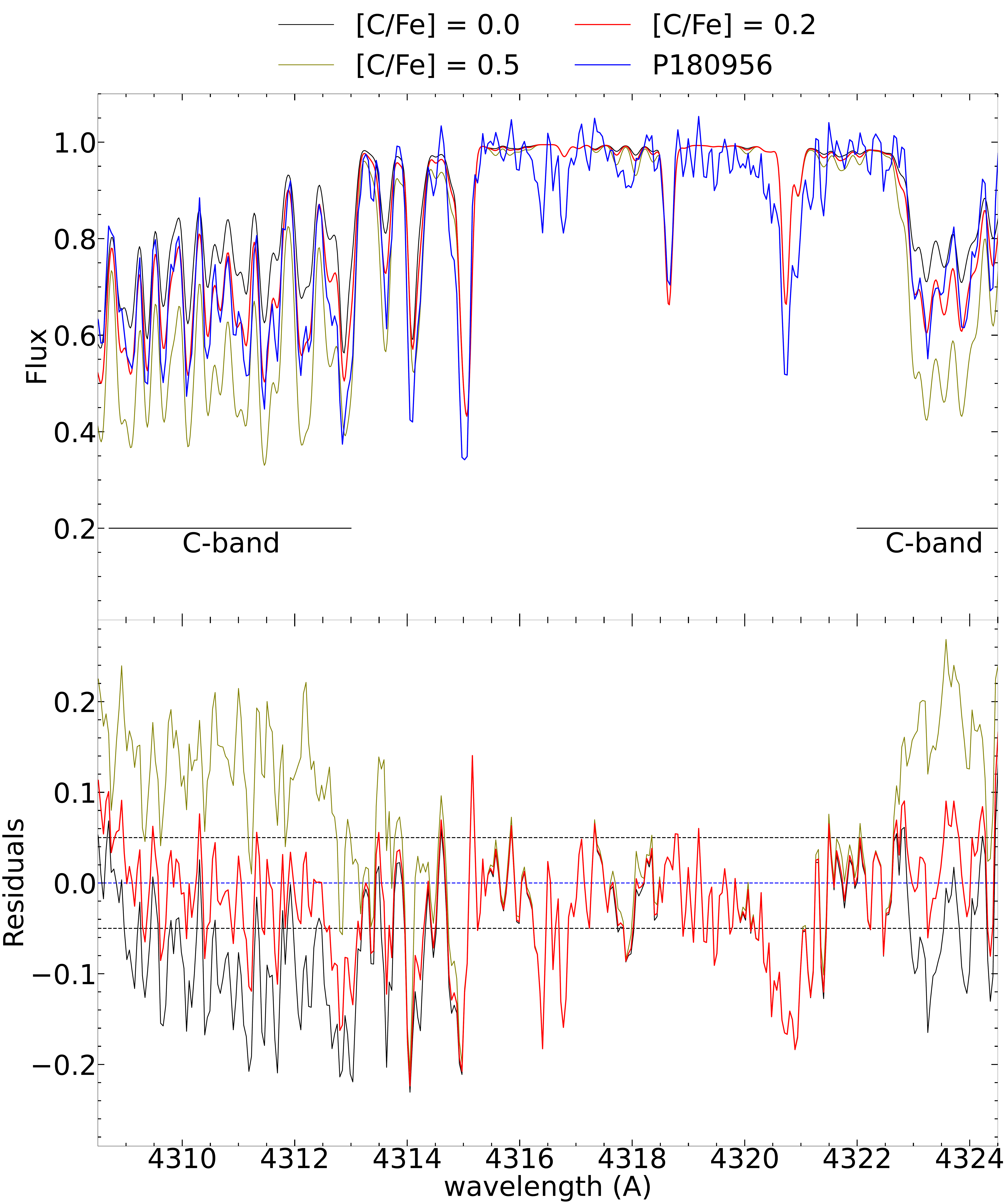}
\caption{Carbon synthesis. Top panel: The observed spectrum of P180956 is marked with a blue line, while synthetic spectra are denoted by  black ([C/Fe] $=0.0$), red ([C/Fe] $=0.2$, best fit), and olive ([C/Fe] $=0.5$) lines. Bottom panel: Residuals of the fits. Horizontal dashed lines marks the null difference (blue) and the dispersion around the continuum ($\pm 0.05$ dex, black). The residuals of the best fit (red line) in the regions of Carbon bands are within or similar to the level of the continuum dispersion.}
\label{Fig:carbon}
\end{figure}

Carbon was first inferred from the low-/medium-resolution campaign of PIGS \citep{Arentsen20a}, showing the star is Carbon-normal, [C/Fe] $=0.17 \pm 0.24$. Figure~\ref{Fig:carbon} shows the synthesis on the CH bands \citep{Masseron14} in the $4300$ \AA{} region (top panel) and the residuals (bottom panel). The synthesis is made with \textsc{synth} mode of \textsc{MOOG}, yielding a [C/Fe] $=0.2 \pm 0.1$, in agreement with the previous measurement. The evolutionary correction for [C/Fe] is inferred using the relation from \citet{Placco14}, which is $0.13$ dex, providing a final  [C/Fe] $=0.33 \pm 0.10$.

\subsection{$\alpha$-elements}
The $\alpha$-elements  producing detectable lines in this star are Mg, Si, Ca, and Ti \citep{Lawler13,Wood13,DenHartog21,DenHartog23,NIST_ASD}. The A(Mg\ione{}) is from two lines of the Mg\ione{} Triplet ($\lambda\lambda 5172.684, 5183.604$\AA) and from other 5 lines for which the SNR is high ($>15$). Si\ione{} is detected from the $4102.936$ \AA{} line, which is blended with the wing of the broader H$\delta$ line. The A(Ca\ione{}) is inferred from 17 spectral lines, from 4200 \AA{} to 6500 \AA. The Ca Triplet has been excluded since it shows strong lines ($>140$ m\AA). Ti\ione{} and Ti\ii{} are present with 17 and 30 lines, respectively.

\subsection{Odd-Z elements}
Four odd-Z elements are detectable in the GHOST spectrum of this object, Na, Al, K and Sc \citep{Lawler19,Roederer21,NIST_ASD}.  Na\ione{}  is present with the Na\ione{} Doublet ($\lambda\lambda 5889.951,5895.924$ \AA). The Na\ione{} D lines from the interstellar medium are not affecting the stellar component,  in agreement with \citet{Sestito23}. The combination of the wide spectral coverage of GHOST and its efficiency in the blue region of the spectrum allows us to detect the 2 lines of Al\ione{} $\lambda\lambda 3944.006,3961.520$ \AA{}. K\ione{} is observable with two lines at $\lambda\lambda7664.899, 7698.965$ \AA{}, which are usually close to water vapour lines of the Earth's atmosphere. In this case, the K\ione{} spectral lines are well separated from the telluric lines. Sc is present with 6 Sc\ii{} lines from 4200 to 5700 \AA{}.

\subsection{Fe-peak elements}
The Fe-peak elements that are observable in the GHOST spectra of this target are Fe, V, Cr, Mn, Co,  Ni,  Cu and Zn \citep{Sobeck07,Melendez09,DenHartog11,DenHartog14,DenHartog19,Roederer12,Ruffoni14,Lawler14,Lawler15,Lawler17,Belmonte17,Wood14,Wood18}.
A(Fe\ione{}) and A(Fe\ii{}) are measured from 188 and 15 lines, respectively. V\ione{} is present with the 4379.230 \AA{} line. Cr\ione{} and Cr\ii{} are detected from 8 and 1 line (4558.650 \AA), respectively. A(Mn\ione{}) is from 7 lines between 4000 to 4850 \AA. A(Co\ione{}) is measured from 4 lines in the blue from 3800 to 4150 \AA. Ni\ione{} is detected from the  $\lambda\lambda 5476.904, 6643.630$ \AA{} lines. An upper limit on A(Cu\ione{}) is obtained with the synthesis on the  $\lambda\lambda 5105.541, 5782.132$ \AA{} lines,  [Cu/Fe] $<-0.2$. Only  line 4810.528 \AA{} of Zn\ione{} is present in this GHOST spectrum.

\subsection{Neutron-capture process elements}
\begin{figure}
\includegraphics[width=0.5\textwidth]{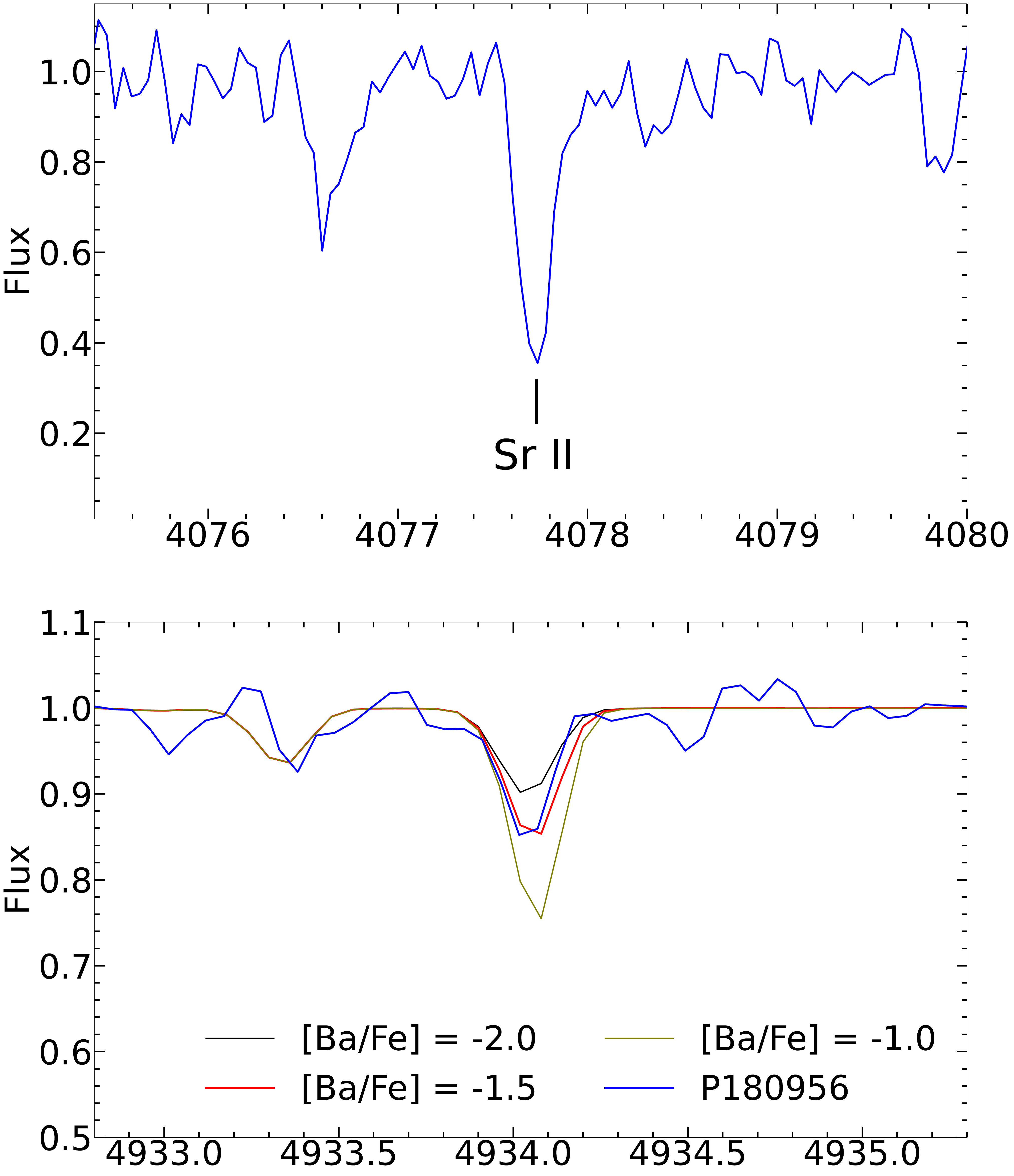}
\caption{Sr and Ba lines. Top panel: one of the Sr lines observed in the GHOST spectrum of P180956. Bottom panel: One of the Ba lines observed is marked with a blue line, while synthetic spectra are denoted by  black ([Ba/Fe] $=-2.0$), red ([Ba/Fe] $=-1.5$, best fit), and olive ([Ba/Fe] $=-1.0$) lines.}
\label{Fig:basr}
\end{figure}

The blue coverage of the GHOST instrument is essential to detect the neutron-capture process elements in the spectrum of this very metal-poor stars, namely Sr, Y, Ba, La, and Eu \citep{Hannaford82,Lawler01,Lawler01b,Biemont11,NIST_ASD}. Sr\ii{} is observed from the $\lambda\lambda 4077.709, 4215.519 $ lines and it is shown in the top panel of Figure~\ref{Fig:basr}. A(Y\ii{}) is measured with the \textsc{synth} mode in \textsc{MOOG} from the $\lambda\lambda 4398.010, 4900.119, 4883.682, 5200.409$ \AA{}, giving an upper limit of [Y/Fe] $<0.0$.  The synthesis of Ba\ii{} is made from 4 lines, $\lambda\lambda 4554.029, 4934.076,6141.730,6496.910 $ \AA{}, which provide a value in agreement with the GRACES analysis, [Ba/Fe] $\sim-1.5$. The observed Ba\ii{} 4934.076 \AA{} line and three synthetic spectra are shown in the bottom panel of Figure~\ref{Fig:basr}. La\ii{} produces very weak lines, $\lambda\lambda 4920.980, 5290.820, 5301.970, 5303.530$ \AA{}, resulting in a high upper limit, [La/Fe] $<0.5$. The Eu\ii{} lines, $\lambda\lambda 4129.700, 4205.050 $ \AA{}, are hard to detect, giving an upper limit of  [Eu/Fe] $<0.0$. Other species of neutron-capture elements have been inspected, which present lines at the level of the noise, resulting in uninformative and very high upper limits.

\subsection{Non-Local Thermodynamic Equilibrium corrections}
The atmosphere of VMP stars are affected by non-local thermodynamic equilibrium (NLTE) effects, which can be large for some species.  NLTE corrections have been applied to Fe\ione{} and Fe\ii{} \citep{Bergemann2012}, Mg\ione{} \citep{Bergemann2017}, Si\ione{} \citep{Bergemann2013}, Ca\ione{} \citep{Mashonkina17}, Ti\ione{} and Ti\ii{} \citep{Bergemann2011}, Cr\ione{} \citep{Bergemann2010b}, Mn\ione{} \citep{Bergemann2019}, and Co\ione{} \citep{Bergemann2010} using the MPIA data base\footnote{\url{https://nlte.mpia.de}}. For Na\ione{} \citep{Lind2012} and Sr\ii{} \citep{Bergemann2012b}, the  \textsc{INSPECT}\footnote{\url{http://www.inspect-stars.com}} webtool was used. Note that NLTE corrections for Sr\ii{} are not available for a star with the stellar parameters of P180956, therefore, the closest parameters logg $=2.3$ (vs. $1.87\pm 0.10$) and v$_{\rm{micro}} =1.0\kms$ (vs. $1.5\pm 0.1\kms$) provide a negligible correction of $\sim-0.01$ dex. Similarly, Ba\ii{} NLTE corrections, adopted from \citet{Mashonkina19}, are not available for this star. The closest match in their online database\footnote{\url{http://www.inasan.ru/~lima/pristine/ba2/}} gives a minor correction of $\lesssim0.05$.  NLTE corrections for K are obtained from \citet{Ivanova00}, which also include hyperfine structure corrections. We highlight that NLTE corrections on Fe, Ti, and Cr are helpful to obtain the ionisation balance among these species, \ie A(X\ione{}) $\approx$ A(X\ii).
 
Table~\ref{tab:chems} reports the chemical abundances ratios in LTE, [X/H]$_{\rm{LTE}}$ and [X/Fe]$_{\rm{LTE}}$, their uncertainties, the number of lines used, and the average NLTE corrections, $\Delta_{\rm{NLTE}}$.

\begin{table}
\caption[]{Chemical abundances [X/H] in LTE, their final uncertainties $\sigma_{\rm{[X/H]}}$, already divided by the square root of the number of lines, the [X/Fe] ratios in LTE, the number of spectral lines used N$_{\rm{lines}}$, and the NLTE corrections $\Delta_{\rm{NLTE}}= \rm{[X/H]}_{\rm{NLTE}} - \rm{[X/H]}_{\rm{LTE}}$. Upper limits are expressed within $1\sigma$.}
\centering
\resizebox{0.47\textwidth}{!}{
\hspace{-0.6cm}
\begin{tabular}{lrrrrr}
\hline
Species & [X/H]$_{\rm{LTE}}$ & $\sigma_{\rm{[X/H]}}$ & [X/Fe]$_{\rm{LTE}}$ & N$_{\rm{lines}}$   & $\Delta_{\rm{NLTE}}$ \\
\hline
Fe\ione{} &   $-1.95$   &  $0.02$ & $-$ &  188 & $+0.148$   \\
Fe\ii{} &   $-1.64$  &  $0.04$ & $-$ & 15  &  $+0.004$  \\
C &   $-1.75$  &  $0.10$ & $+0.20$  &  $-$  &  $-$  \\
Na\ione{} &   $-2.38$  &  $0.05$ & $-0.43$ & 2  &  $-0.457$  \\
Mg\ione{} &   $-1.94$  &  $0.09$ & $+0.01$ & 7  &   $+0.089$ \\
Al\ione{} &   $-2.20$  &  $0.25$ & $-0.25$ & 2  & $-$   \\
Si\ione{} &   $-1.25$  &  $0.10$ & $+0.70$ &  1 &  $+0.040$  \\
K\ione{} &    $-1.40$ &   $0.13$ & $+0.55$ & 2  &  $-0.300$  \\
Ca\ione{} &   $-2.03$  &  $0.05$ & $-0.08$ & 17  &  $+0.173$  \\
Sc\ii{} &   $-1.23$  &  $0.07$ & $+0.72$ & 6  &  $-$  \\
Ti\ione{} &   $-1.81$  &  $0.04$ & $+0.14$ & 17  &  $+0.509$  \\
Ti\ii{} &   $-1.27$  &  $0.03$ & $+0.68$ & 30  &  $-0.038$  \\
V\ione{} &    $-2.07$ &   $0.10$ & $-0.12$ & 1  &  $-$  \\
Cr\ione{} &   $-2.23$  &  $0.04$ & $-0.28$ & 8  &  $+0.382$  \\
Cr\ii{} &   $-1.87$  &  $0.10$ & $+0.08$ & 1  &  $-$  \\
Mn\ione{} &   $-2.40$  &  $0.05$ & $-0.45$ &  7 &   $+0.381$ \\
Co\ione{} &   $-1.59$  &  $0.21$ & $+0.36$ & 4  &  $+0.583$  \\
Ni\ione{} &   $-1.82$  &  $0.04$ & $+0.13$ & 2  &  $-$  \\
Cu\ione{} &   $<-2.20$  &  $-$ & $<-0.25$ & 2  &  $-$  \\
Zn\ione{} &   $-1.66$  &  $0.10$ & $+0.29$ & 1  &   $-$  \\
Sr\ii{} &   $-2.25$  &  $0.01$ & $-0.30$ & 2  &  $-0.011$  \\
Y \ii{} &   $<-2.00$  &  $-$ & $<-0.05$  & 4  & $-$   \\
Ba\ii{} &   $-3.50$  &  $0.10$ & $-1.55$ &  4 &  $<0.05$  \\
La\ii{} &   $<-1.50$  &  $-$ & $<+0.45$ &  4 &  $-$  \\
Eu\ii{} &   $<-2.00$  &  $-$ & $<-0.05$ & 2  &  $-$  \\
\hline
\end{tabular}}
\label{tab:chems}
\end{table}

\subsection{Comparison with the GRACES spectral analysis}\label{sec:comparisonGRACES}
One of the reasons that this star P180956 was selected as a GHOST commissioning target was its unusually low [Na/Mg], [Ca/Mg], and [Ba/Fe] ratios, either in LTE or NLTE \citep{Sestito23}.  Thus, we examine our abundance results from this improved GHOST spectrum in comparison to those from the more limited GRACES spectrum.

\begin{figure}
\includegraphics[width=0.5\textwidth]{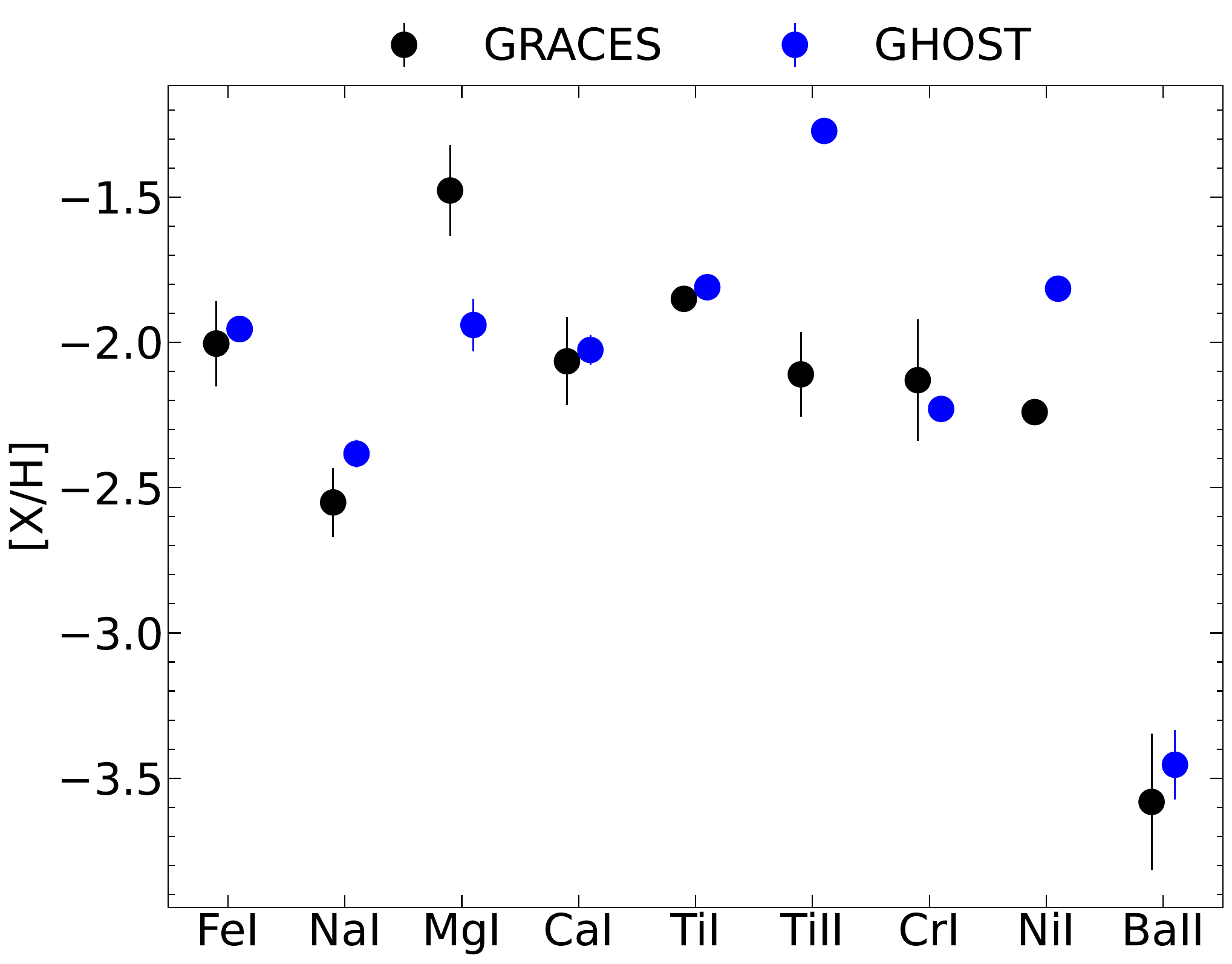}
\caption{GHOST vs. GRACES chemical abundances, [X/H]. GRACES abundances are from \citet{Sestito23}. Both datasets are in LTE. Points are slightly offset in the horizontal direction to better show their errorbars. Uncertainties on GHOST data are often smaller than the size of the marker.
}
\label{Fig:diff}
\end{figure}

In general, the chemical abundance results from the GHOST spectra are similar to those from the previous GRACES analysis, as seen in Figure~\ref{Fig:diff}, both in LTE. The agreement is excellent for the majority of the elements ($\lesssim2\sigma$), as expected given that we have adopted the same stellar parameters. The better quality of the GHOST spectrum at all  wavelengths and the larger  number of lines produce a smaller line-to-line dispersion, and, therefore,  smaller uncertainties on [X/H].

Our Fe\ione{} abundance from GHOST is determined from 188 lines, whereas the GRACES analysis included only 63 lines, which improves the line-to-line scatter  by a factor of $\sim1.5$. 

Larger differences ($\gtrsim 3\sigma$) are found for Mg\ione{}, Ti\ii{}, and Ni\ione{}. Looking at the previous analysis \citep{Sestito23}, this star has been analysed setting a lower metallicity value in the model atmosphere then the final one (by $\sim0.4$ dex)\footnote{We reassure the reader that this is the only star in \citet{Sestito23} for which the model atmosphere was not updated to the final value}. On the other hand, EW measurements of lines in common to both analyses are similar, to within $\sim5-10$ percent. Hence, the former should be the main culprit to the differences we found for Mg\ione{}, Ti\ii{}, and Ni\ione{}.

\subsection{Comparisons with the MW halo and bulge}
\begin{figure*}
\includegraphics[width=1\textwidth]{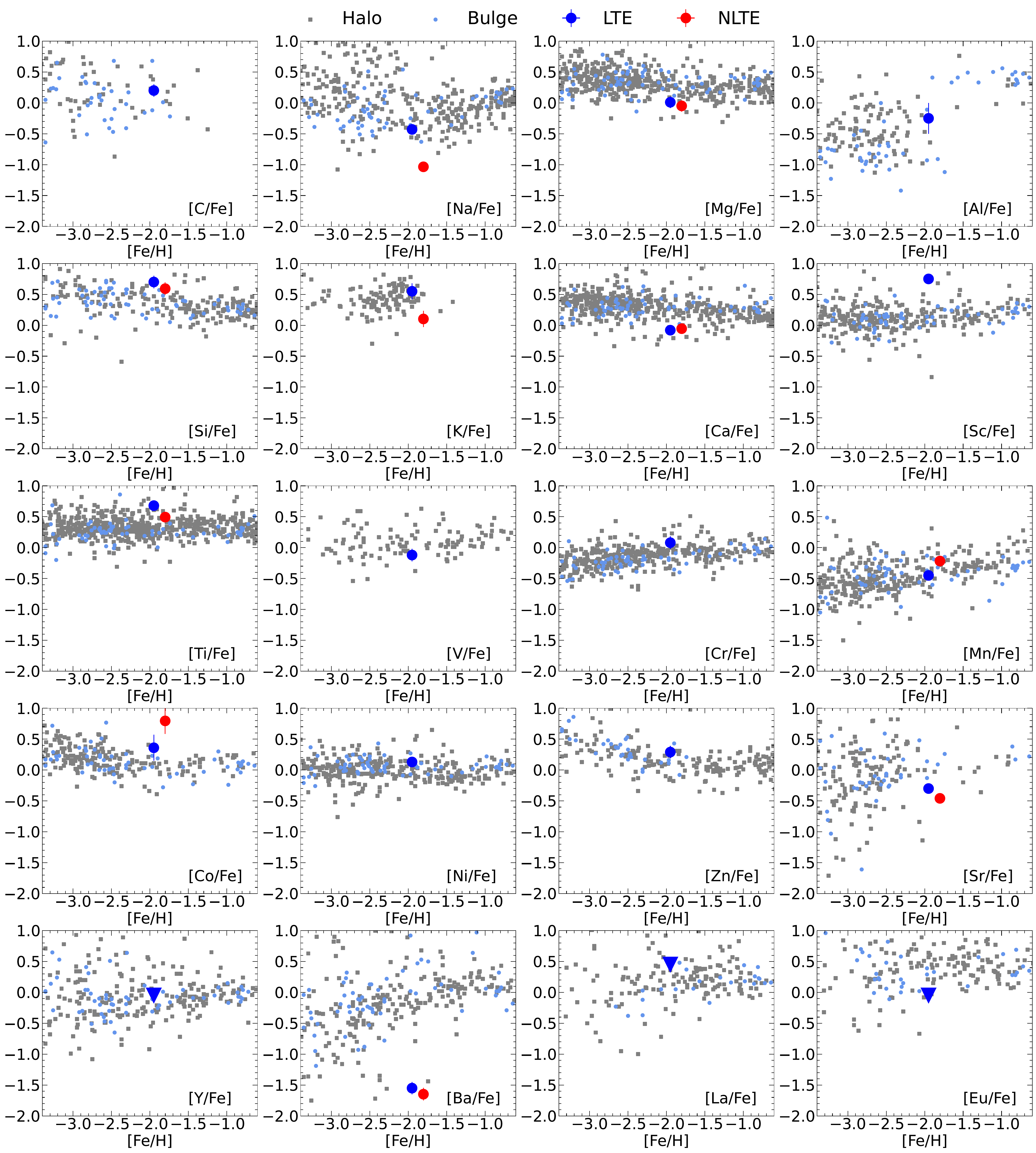}
\caption{Chemical abundances. Blue and red circles represent the chemical abundances of P180956 in LTE and NLTE-corrected, respectively. Blue triangles denote upper limits for P180956. Small light blue circles represent the bulge compilation \citep{Howes14,Howes15,Howes16,Koch16,Reggiani20,Lucey22}, while small grey squares correspond to the halo compilation from the SAGA database \citep{Suda08}. The uncertainties on [X/Fe] for P180956 are often smaller than the symbol size. The scarcity of literature stars in the [K/Fe] panel may be attributed to the difficulty in measuring K spectral lines in the optical range, as they can be blended with telluric water vapour lines.
}
\label{Fig:chems}
\end{figure*}

The chemical abundances of P180956 are compared to a compilation of stars in the MW bulge and halo in Figure~\ref{Fig:chems}. The panels in the figure are arranged in order of increasing proton number of the species. The MW bulge compilation (small light blue circles) includes results from \citet{Howes14,Howes15,Howes16}, \citet{Koch16}, \citet{Reggiani20}, and \citet{Lucey22}. The MW halo compilation (small grey squares) consists of stars obtained from the Stellar Abundances for Galactic Archaeology database\footnote{\url{http://sagadatabase.jp}} \citep[SAGA,][]{Suda08}, with high-resolution analysis (R~$>30000$),  no lower or upper limits on the measurements, and low uncertainties on the chemical abundances, $\sigma_{\rm{[X/H]}}<0.2$. Both the compilations are from LTE analyses.

The chemical ratios of [Na, Mg, Ca, Sr, Ba/Fe] and upper limit for [Eu/Fe] in P180956 are situated at the lower end of the distribution observed in the MW halo and bulge. This is particularly evident for [Ba/Fe], which is nearly 2 dex lower than the majority of stars at the same \FeH. In contrast, [Si, Ti, Sc, Co/Fe] ratios in P180956 are slightly enhanced compared to the literature compilations for the MW. Other upper limits (Cu, Y, La) do not provide significant constraints on the abundances.

In the GRACES analysis \citep{Sestito23}, P180956 had  very low [Na, Ca/Mg] ratios. Here, the revised ratios are [Na/Mg]$_{\rm{NLTE}} = -0.99 \pm 0.10$ and [Ca/Mg]$_{\rm{NLTE}} = -0.01 \pm 0.10$. The latter is now in line with MW halo and dwarf galaxies expectations \citep[\eg][]{Hansen20,Ji20}. Comparing the other $\alpha-$elements to Mg, this star shows [Si/Mg]$_{\rm{NLTE}} = +0.64 \pm 0.10$ and [Ti/Mg]$_{\rm{NLTE}} = +0.55 \pm 0.10$. Thus, P180956 exhibits very low [Na/Fe], slightly sub-solar [Mg, Ca/Fe],  enhanced [Si, Ti/Mg], in addition to a very low Ba-content.  This is a somewhat unusual chemical abundance pattern for a typical MW halo or bulge star.

\section{Galactic Orbit}\label{sec:orbit}
\begin{figure*}
\includegraphics[width=1\textwidth]{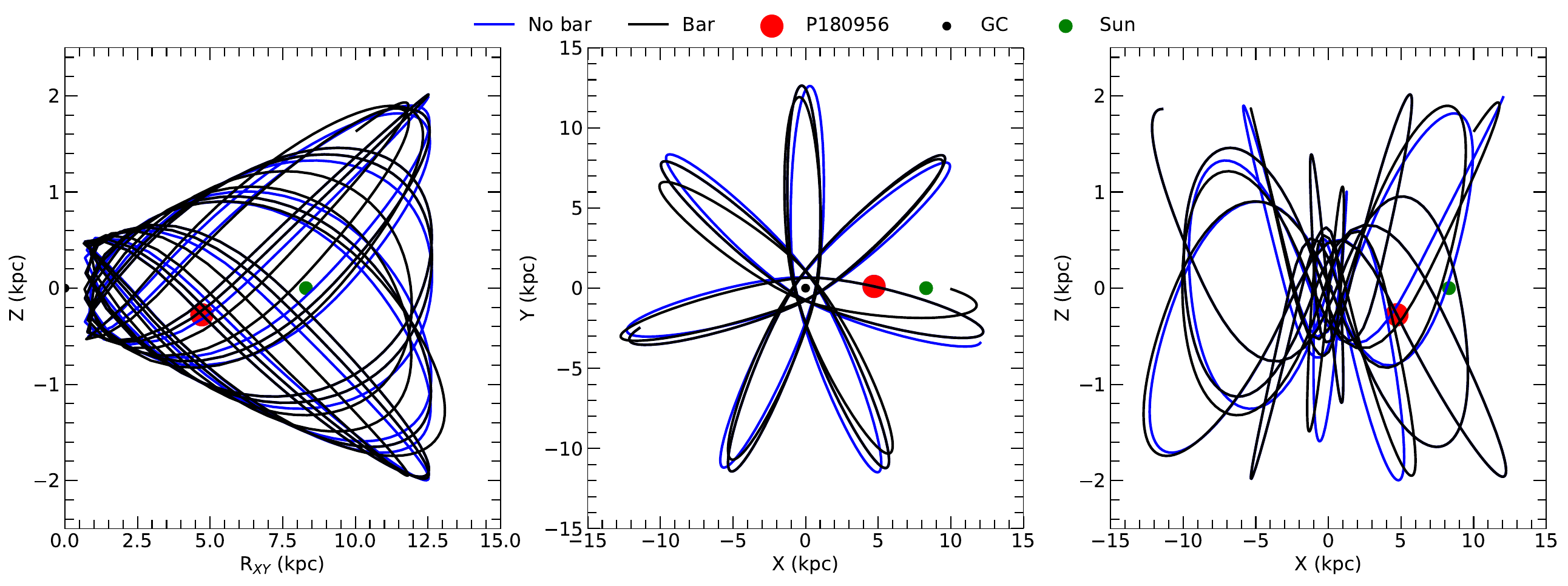}
\caption{Galactic orbit of P180956. Left panel: Height from the plane Z vs. in-plane projected distance R$_{\rm{XY}}$. Central panel: Y vs. X Galactic positions. Right panel:  Z vs. X Galactic positions. Black and blue lines represent the orbit integrated in a gravitational potential with and without the presence of a rotating bar as in \citet{Sestito23} and as in \citet{Sestito19}, respectively. The red, black and green circles mark the position of P180956, of the Galactic centre (GC), and of the Sun at the present day. 
}
\label{Fig:orbit}
\end{figure*}

Orbital parameters are derived using \textsc{Galpy} code \citep{Bovy15}, integrating the orbit for 1 Gyr in the future and in the past. Uncertainties are determined through  a Monte Carlo simulation (1000 iterations) on the input quantities, drawn from a Gaussian distribution. Two Galactic gravitational potentials are adopted and the relative orbits are displayed in Figure~\ref{Fig:orbit}. One potential corresponds to the model used in \citet{Sestito23}, which includes a rotating bar (black line); the second (blue line) is without the bar as in \citet{Sestito19}. These two gravitational potentials lead to a very similar orbit.

In both cases, P180956 exhibits a slightly prograde orbit (vertical angular momentum L$_{\rm{z}} \sim 300$ kpc $\kms$), which is confined to the Milky Way plane (Z$_{\rm{max}} \sim 2\kpc$). The pericentric (R$_{\rm{peri}} \sim 0.7 \kpc$) and apocentric (R$_{\rm{apo}} \sim 13 \kpc$) distances indicate that the star's trajectory takes it very close to the Galactic centre before venturing far beyond the Sun's position, resulting in an orbit characterised by high eccentricity ($\epsilon \sim 0.9$). These results are in agreement  within $1\sigma$ with the previous orbital analysis from \citet{Sestito23}.

Table~\ref{Tab:orbit} reports pericentric and apocentric distances, the eccentricity, the maximum excursion from the plane, the vertical component of the angular momentum as calculated using both gravitational potentials. 

\begin{table}
\caption[]{Pericentric and apocentric distances (R$_{\rm{peri}}$, R$_{\rm{apo}}$), the eccentricity ($\epsilon$), the maximum excursion from the plane (Z$_{\rm{max}}$), and the vertical component of the angular momentum (L$_{\rm{z}}$) are reported.}
\centering
\resizebox{0.47\textwidth}{!}{
\hspace{-0.6cm}
\begin{tabular}{lcc}
\hline
Quantity & Bar & No Bar \\
\hline
R$_{\rm{peri}}$ (kpc) &  $0.67  \pm  0.21 $ &  $0.73  \pm  0.18$ \\
R$_{\rm{apo}}$ (kpc)  & $12.82 \pm  1.06$& $12.67  \pm  0.74$ \\
$\epsilon$  & $0.90  \pm  0.03$ & $ 0.89  \pm  0.02$ \\
Z$_{\rm{max}}$ (kpc)  & $2.01  \pm  0.35$ & $2.02  \pm  0.33 $\\
L$_{\rm{z}}$ (kpc $\kms$) & $307.0  \pm  83.0$ &  $303.0  \pm  85.0$ \\
\hline
\end{tabular}}
\label{Tab:orbit}
\end{table}

\section{Discussion}\label{sec:discussion}
P180956 has peculiar kinematics that indicate that the star is confined to the Milky Way plane on a very eccentric orbit, reaching both the very inner region of the MW and a position well beyond the Sun (see Figure~\ref{Fig:orbit}). 

Furthermore, P180956 showed a very low amount of [Ba/Fe], [Ca/Mg], and [Na/Mg]. These signatures stand out when compared to MW halo stars, and  resemble the abundances of stars found at present in ultra-faint dwarf galaxies.  Low chemical abundance ratios in the elements listed above have been interpreted as a sign of contributions from a small number of low-mass supernovae type II (SNe~II) in the past, \eg the so-called "one-shot" model \citep{Frebel10}. 
Therefore, the interpretation of the  chemo-dynamical properties from the GRACES spectrum implied that this star may have formed in an ancient dwarf galaxy accreted very early in the MW formation. In the following subsections, a revised and more thorough discussion on the origin of this star and on the properties of its formation site is presented.

\subsection{The yields of the supernovae progenitors}\label{sec:yields}
The wavelength coverage of GHOST and the SNR of the observed spectra allow the detection of up to 18 chemical species and four meaningful upper limits, for a total of 13  elements beyond those  available in the previous GRACES spectral analysis. Comparing this extensive set of chemical abundances with the predicted yields of supernovae from theoretical models can be used for insights into the nucleosynthetic processes that occurred at the formation site of P180956. To accomplish this, the online tool \textsc{StarFit}\footnote{\url{https://starfit.org}} is used. The yields of best fit are obtained by combining the type II supernovae yields of the best solutions chosen from a pool of theoretical models. A total of ten models have been selected, encompassing various types of supernovae events, \ie hypernovae, core-collapse, rotating massive stars, neutron stars mergers, and pair-instability SNe. 

\begin{figure}
\includegraphics[width=0.5\textwidth]{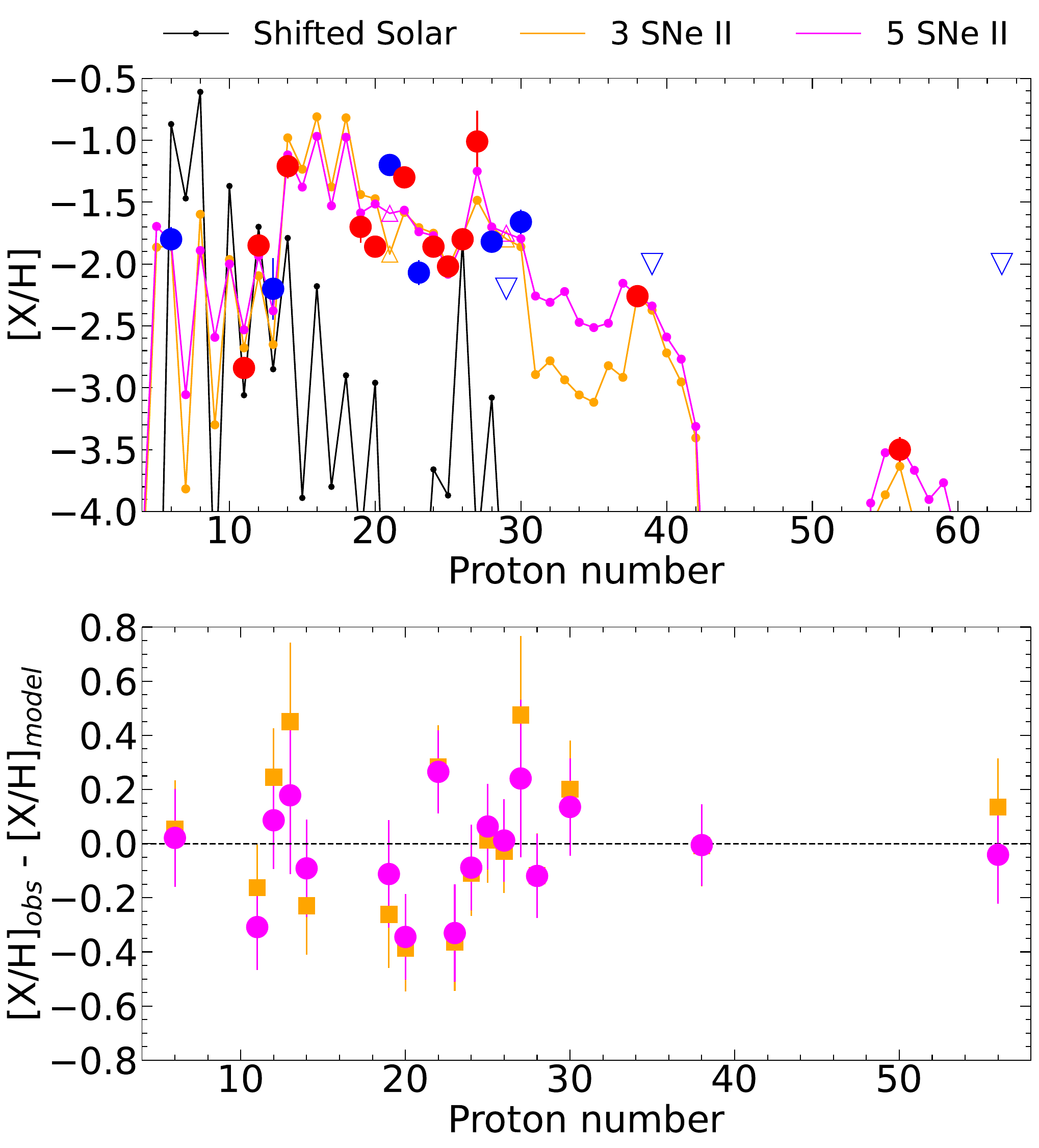}
\caption{Supernovae yields. Red and blue circles are NLTE and LTE [X/H] of P180956, respectively. Upper limits of Cu, Y, and Eu are marked with empty downward blue triangles. Top panel: Black dots and line represent the solar scaled abundances from \citet{Asplund09} shifted to match the \FeH{} of P180956. Magenta and orange dots and lines represent the theoretical yields from the best (5 SNe~II events) and second best fit (3 SNe~II events), respectively. Lower limits of Sc and Cu in the theoretical models are denoted with empty upward triangles. Bottom panel: The difference between the observed [X/H] and the yields predicted by the best (magenta circles) and second best (orange squares) fit from \textsc{StarFit}. Upper limits in the data  and lower limits in the modelled yields are removed. Uncertainties on the data are summed in quadrature to  $0.15$ for the theoretical yields. The horizontal dashed line marks the null difference.}
\label{Fig:yields}
\end{figure}

The theoretical yields [X/H] from contributing SNe are compared with the observational data in the top panel of Figure~\ref{Fig:yields}. The scaled solar abundances ratios (black line) are from \citet{Asplund09}, shifted to match the \FeH{} of P180956.  These fail to reproduce our data, except for Fe (by design), Na, and Mg. The solar abundances pattern predicts a net decrease in the yields from Si, which is not seen in P180956.

The best fit solution (magenta line)  from \textsc{StarFit} consists in five SNe~II events originating from Population III stars. This fit is a mixture of three low-mass hypernovae\footnote{A hypernova explosion has a typical energy of at least a factor $\sim10$ greater than a classical supernova type II.} with M $\sim10\msun$ \citep{Heger10,Heger12}, one around M $\sim15\msun$ \citep{Limongi18}, and one intermediate-mass ($\sim80\msun$) fast-rotating ($\sim300\kms$) supernova  \citep{Grimmett18}. The second best fit (orange line) solution is composed of three events, two low-mass hypernovae with M $\sim10-17\msun$ \citep{Heger10,Heger12} and one fast-rotating ($\sim300\kms$) SN with M $\sim120\msun$ \citep{Limongi18}. 

The difference between the observed [X/H] and the theoretical yields from the best (magenta circles) and second best (orange squares) fits are displayed in the bottom panel of Figure~\ref{Fig:yields}. Both solutions provide a difference below $\lesssim0.3$  in absolute value for the majority of the species. 

We want to emphasise that the use of \textsc{StarFit} is more as an illustration of the range of events that might be needed to explain the chemistry of this star. \textsc{StarFit} suggests an interpretation that contributions from high-mass supernovae ($>140\msun$, \eg pair-instability) and neutron-star mergers are ruled out for this star. The former would produce a strong odd-even effect in the yields \citep{Takahashi18,Salvadori19}, \ie very low [Na, Al/Mg] ($\sim-1.3$) and high [Ca/Mg] ($\gtrsim +0.6$); the latter would produce an enrichment in neutron-capture elements \citep[\eg][]{Cowan21}. Both scenarios are in contrast with the observed chemical properties of P180956. \textsc{StarFit} suggests that the main contributing  SNe~II are in the low-mass range. Specifically, hypernovae are necessary to produce the high [Si, Ti/Mg] ratios,  and they provide a little contribution to heavy elements. These findings align with the scenario proposed by \citet{Ishigaki18}, where low-mass hypernovae ($\lesssim40\msun$) are the primary sources of enrichment of the interstellar medium during the early stages of chemical evolution. In addition, the presence of one fast-rotating intermediate-mass supernova is needed to well reproduce the pattern of the heavy elements, \ie Sr, Ba.

\subsection{In-situ vs. accreted diagnostics}\label{sec:inexsitu}
What if this star formed in-situ, \ie in the proto-disc but after the early Galactic assembly?  The period between the early MW assembly and the formation of the disc, dubbed "Aurora" \citep{Belokurov22}, has been proposed to be very chaotic, forming bound massive clusters, chemically similar to globular clusters \citep{Belokurov23}. This implies that some Aurora stars would be enriched in  N, Na, and Al, while others would resemble the "normal" halo stars \citep{Belokurov23}. P180956 is Na-poor, Al-normal, and neutron-capture process poor, with the latter being very rare for halo stars at that metallicity (see Figure~\ref{Fig:chems}). This chemical pattern rules out the "Aurora" as the origin for P180956. 

\begin{figure}
\includegraphics[width=0.5\textwidth]{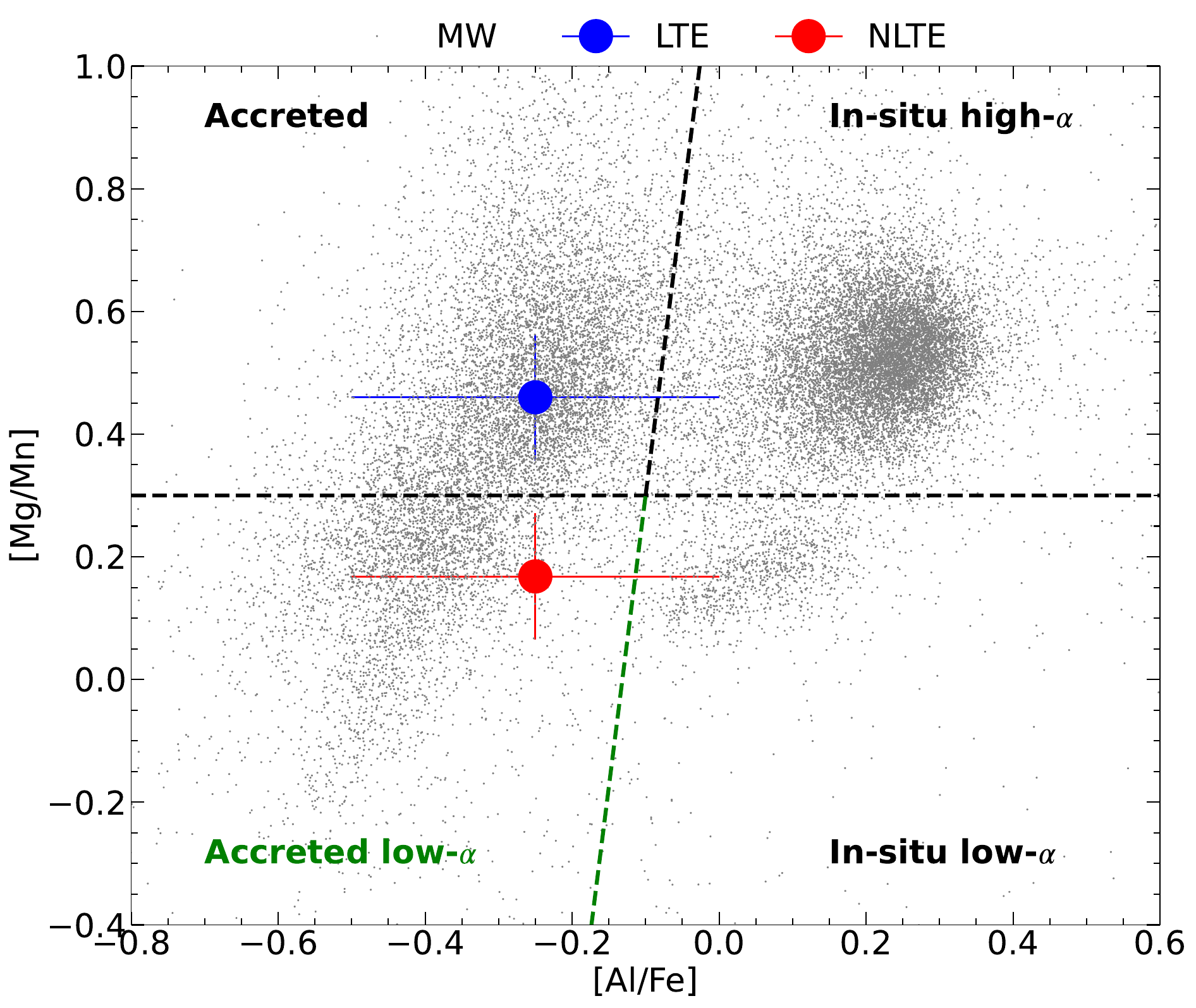}
\caption{[Mg/Mn] vs. [Al/Fe] space. P180956 data are shown in blue and red in LTE and NLTE-corrected ([Mg/Mn] only),  respectively. Small grey dots are APOGEE DR17 MW stars selected to have SNR $>70$ and $\FeH<-0.7$, which are derived from an NLTE analysis. The three regions delimited by the dashed black lines, accreted and in-situ low-/high-$\alpha$ are defined following \citet{Das20} and \citet{Horta21}. The accreted region has been prolonged following the dashed green line, in which accreted low-$\alpha$ stars should lie.}
\label{Fig:mgmn}
\end{figure}

Other studies have suggested [Mg/Mn] vs. [Al/Fe] as a diagnostic to differentiate accreted stars from in-situ stars \citep[\eg][]{Das20,Horta21}. Figure~\ref{Fig:mgmn} illustrates this chemical space, including  APOGEE DR17 \citep{APOGEEDR17} stars for comparison. Dashed black lines delineate three regions where accreted, in-situ low-$\alpha$, and in-situ high-$\alpha$ stars are more likely to be found \citep[\eg][]{Das20,Horta21}. In the LTE case, P180956 lies close to the centre of the accreted "blob", while considering NLTE corrections, the star falls into the in-situ low-$\alpha$ region.

Recently, \citet{Horta23} provide a summary of the chemo-dynamical properties of  known accreted structures based on the latest Gaia and APOGEE data releases. Their Figure~13 displays the [Mg/Mn] vs. [Al/Fe] space for all the accreted known structures, showing that some of their stars have [Mg/Mn] $\sim0.0$ or even negative (\eg Sagittarius, Sequoia, Gaia-Sausage-Enceladus). Therefore, part of the in-situ low-$\alpha$ region, is actually accreted. The accreted region in Figure~\ref{Fig:mgmn} is here tentatively extended (accreted low-$\alpha$), by lengthening the dashed green line, which suggests that P180956 is of accreted origin. Offsets between the NLTE infra-red (APOGEE) and the NLTE-corrected optical analyses are estimated to be $\Delta\rm{[Mg/Mn]}= +0.15\pm 0.18$ and $\Delta\rm{[Al/Fe]}= -0.11 \pm 0.12$ \citep{Jonsson20}. These corrections, not applied in Figure~\ref{Fig:mgmn}, would move P180956 more closely to the accreted low-$\alpha$ "blob".

Can P180956 be associated with any of the recently discovered accreted structures?  The system that bears the closest resemblance to P180956 in terms of kinematical properties \citep[see][]{Horta23} is Gaia-Sausage-Enceladus \citep[GSE, \eg][]{Helmi18,Belokurov18}. The high eccentric orbit of GSE ($0.93\pm0.06$) and its mean pericentric and apocentric distances ($r_{\rm{peri}} = 0.61 \pm 1.03\kpc$ and $r_{\rm{apo}} =17.15 \pm 5.22\kpc$) are compatible with P180956's orbital parameters within $1\sigma$ (see Table~\ref{Tab:orbit}). However, stars of GSE  reach a maximum height from the plane that is typically greater than the one of our target ($Z_{\rm{max}} = 9.84 \pm 6.14 \kpc$ vs. $Z_{\rm{max}} = 2.01 \pm 0.35 \kpc$).

\citet{Helmi18} demonstrate that the GSE population has lower [$\alpha$/Fe] ratios compared to stars in the MW halo. Similarly, \citet{Hayes18} provide evidence for a low-$\alpha$ population, showing that its stars exhibit [Mg, Si/Fe]$_{\rm{NLTE}}\sim 0.25-0.30$ at the \FeH{} of P180956. However, these ratios are not compatible with the values of our target, which have [Mg/Fe]$_{\rm{NLTE}}= -0.05 \pm 0.09$ and [Si/Fe]$_{\rm{NLTE}}= +0.59 \pm 0.10$. Thus, the chemo-dynamical properties of P180956 differ from those of GSE and other known accreted structures.

\subsection{Origin in an ultra-faint dwarf galaxy}\label{sec:ufdcomp}
\begin{figure*}
\includegraphics[width=1\textwidth]{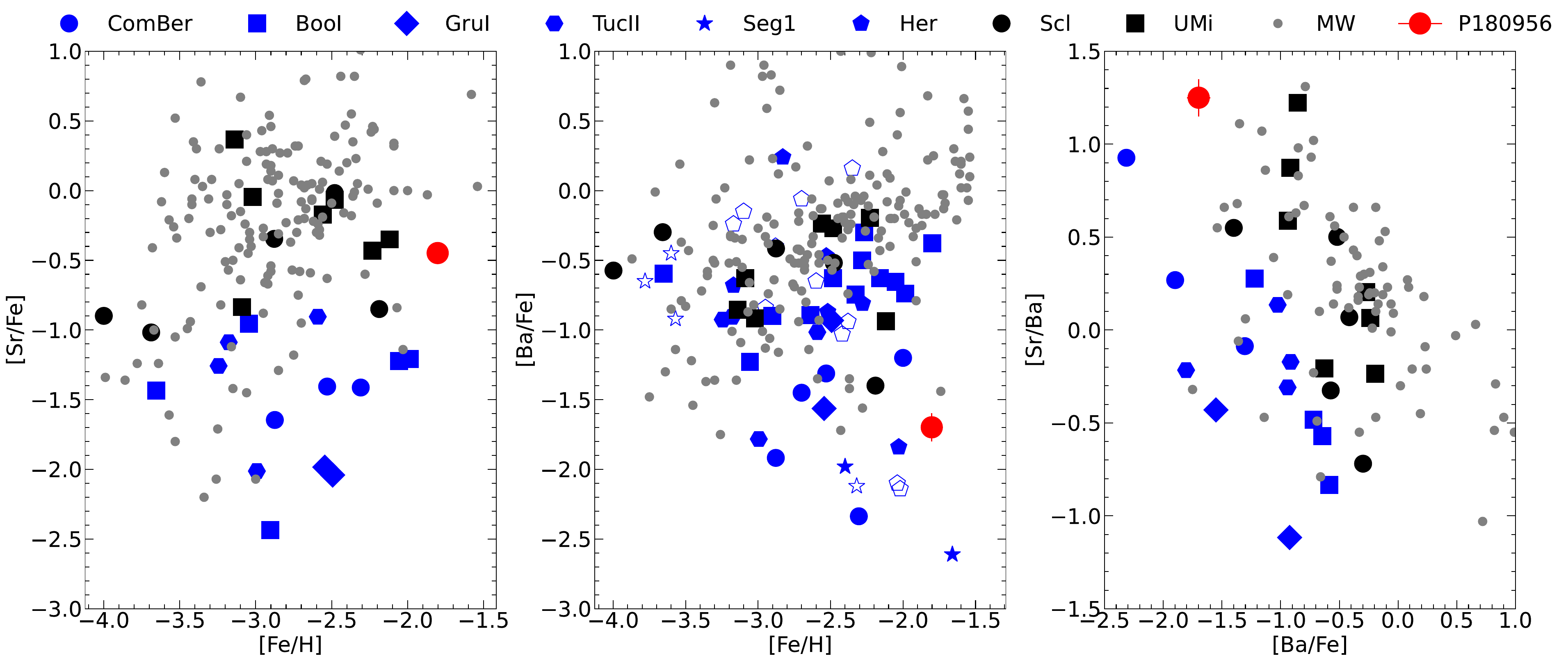}
\caption{Neutron-capture elements. Left panel: [Sr/Fe] vs. \FeH{}. Central panel: [Ba/Fe] vs. \FeH{}. Empty symbols denote upper limits on the vertical axis. Right panel: [Sr/Ba] vs. [Ba/Fe]. The red circle marks the NLTE-corrected chemical ratios of P180956. MW stars are from the SAGA database \citep{Suda08}; Coma Berenice (ComBer) stars are from \citet{Frebel10b} and \citet{Waller23}; Bootes~I (BooI) stars are from \citet{Feltzing09}, \citet{Norris10}, \citet{Gilmore13}, \citet{Ishigaki14}, and \citet{Frebel16}; Gru~I  stars are from \citet{Ji19}; Tucana~II (TucII) stars are from \citet{Ji16} and \citet{Chiti18}; Segue~1 (Seg1)  stars are from \citet{Frebel14}; Hercules (Her) stars are from \citet{Koch08}, \citet{Koch13}, and \citet{Francois16}; Sculptor (Scl) and Ursa Minor (UMi) stars are from \citet{Mashonkina17b}.
}
\label{Fig:neutron}
\end{figure*}

A comparison of the neutron-capture elements [Sr/Fe] and [Ba/Fe] vs. \FeH{} for P180956, stars from ultra-faint dwarf (UFDs), from classical dwarf galaxies (DGs), and Milky Way halo  is presented in the left panels of Figure~\ref{Fig:neutron}. It is evident that P180956 exhibits lower [Sr, Ba/Fe] (and [Eu/Fe] from Figure~\ref{Fig:chems}) ratios than the majority of MW halo stars at the same metallicity. The [Sr/Fe] of P180956 is similar to the bulk value of DGs, while UFDs have lower values. On the opposite behaviour is [Ba/Fe], where P180956's ratio is similar to those in the low-end distribution of UFDs. The low-abundance of neutron-capture processes elements has been interpreted as the contribution of low-mass supernovae and the absence of neutron stars mergers events \citep[\eg][]{Cowan21}. Furthermore, the combinations of stochasticity in the production of neutron-capture elements with the inability of UFDs to retain metals can explain the low [Sr, Ba, Eu/Fe] observed in these systems \citep[\eg][]{Venn12,Ji19}.

The distribution of [Sr/Ba] vs. [Ba/Fe] is shown in the right panel of Figure~\ref{Fig:neutron}.  Halo stars exhibit a downward trend as  [Ba/Fe] increases \citep{Mashonkina17b}, albeit mostly clumped around [Sr/Ba] $\sim 0.3$ \citep{Ji19}. Both UFDs and DGs present a wide distribution in [Sr/Ba] (up to $\sim1.5$ dex), with UFDs populate a distinct region from the majority of  MW halo stars and of DGs \citep{Mashonkina17b,Roederer17,Ji19,Reichert20,Sitnova21}.  P180956 has a relatively high [Sr/Ba] ratio, close to the upper-end of the distributions of  the MW halo, Coma Berenices (UFD), and Sculptor (DG), while its [Ba/Fe] is typical of an UFD's star. 

Did this star originate in an UFD or in classical DG? At the metallicity of P180956,  DGs would likely involve contributions from asymptotic giant branch stars (AGBs) and, in some cases, SNe~Ia \citep[see][for Sculptor and Ursa Minor]{delosreyes20,Sestito23Umi}. AGBs would produce Ba via s-process nucleosynthesis \citep[\eg][]{Pignatari08,Cescutti14}, reaching solar values. Additional contribution from SNe~Ia would lower the overall [$\alpha$/Fe] ratios to solar or sub-solar values. Given P180956 has a very low-Ba ([Ba/Fe]$_{\rm{NLTE}} \sim -1.7$) and the inhomogeneity in the $\alpha$-elements ([Si, Ti/ Mg, Ca] $\sim0.5$), the contributions of AGBs and SNe~Ia are likely ruled out, and, as well, the DG scenario.

How to explain the high [Sr/Ba]? \citet{Mashonkina17b} discuss that sub-solar [Sr/Ba] implies that both  elements are produced solely by r-process, while solar- and supersolar- [Sr/Ba] indicate the involvement of  s-processes in the  Sr production.  Various kinds of supernovae events have been proposed to explain the relative excess of Sr compared to other neutron-capture elements \citep[][and references therein]{Mashonkina17b}. Hypernovae \citep{Izutani09} and  s-process nucleosynthesis in low-metallicity fast-rotating supernovae \citep{Pignatari08,Banerjee18,Grimmett18,Limongi18} are also listed among these, which are always invoked by the best fit models from \textsc{StarFit} (see Section~\ref{sec:yields}). We note from \textsc{StarFit} that the fast-rotating supernova is the event the contribute most to the [Sr/Ba] enrichment.

\subsection{P180956, witness of the early Galactic assembly}
Above we have outlined the most likely origin for P180956 as from an accreted UFD. Two new questions now arise whether the star was brought in during the early Galactic assembly or later on and if its system was isolated or brought in with one of the massive known accreted satellites.

P180956 was also selected for GHOST commissioning because of its peculiar orbital parameters, as it remains relatively close to the MW plane (see Figure~\ref{Fig:orbit}).  Recently, the presence of this population has been observed from the ultra metal-poor regime \citep[UMP, $\FeH\leq-4.0$,][]{Sestito19} to the disc's metallicity \citep{Sestito20,DiMatteo20,Venn20,Cordoni21,Mardini22}, finding that the majority of them moves in a prograde orbit. While VMP "planar" stars have been dynamically detected in various  investigations, thorough and detailed  analyses of their chemical properties are scarce.

The existence of this population has also been explored through high-resolution cosmological zoom-in simulations \citep{Sestito21,Santistevan21}. The  simulations\footnote{\citet{Sestito21} utilised the NIHAO-UHD simulations \citep{Buck19}, while \citet{Santistevan21} employed the FIRE simulations \citep{Hopkins18}.} predict the presence of a "planar" population. While previous observational studies focused on the origin of low-eccentricity stars, \citet{Sestito21} also discussed that the more eccentric members of the planar population are likely brought in during the early-assembly phase. This is because, during that epoch, the gravitational potential of the forming proto-Galaxy is still shallow, allowing for the deposit of accreted systems into the inner regions. The relatively small excursion from the MW plane,  the pericentric and apocentric distances, in addition to the high eccentricity suggest that the star was brought in during the early-assembly phase. 

In Section~\ref{sec:inexsitu}, we rule out that P180956 formed in one of the known accreted structures, however,  given 1) that some of its orbital parameters are similar to those of GSE (except for Z$_{\rm{max}}$), 2) its formation likely in an UFD-like system (see Section~\ref{sec:ufdcomp}), and 3) its early accretion to the MW, two scenarios on the origin of P180956 are proposed. The first is that P180956 originate in one of the many low-mass building blocks that formed the proto-MW, which had chemical and physical properties similar to those of present UFDs; the second is that its progenitor system was an UFD satellite of GSE, that has been brought in into the inner Galaxy during the infall of GSE.

\section{Conclusions}\label{sec:conclusions}
This work represents the second high-resolution spectroscopic analysis utilising the GHOST spectrograph mounted at Gemini South, following the detailed analysis of two r-process rich stars in the Reticulum II ultra-faint dwarf galaxy \citep{Hayes23}. The spectra of P180956, a star with unique chemo-dynamical properties, were observed during the second commissioning run of the instrument in September 2022. Previously, the star was observed with GRACES at Gemini North and analysed as part of the Pristine Inner Galaxy Survey \citep{Arentsen20a,Sestito23}. In this study, we conducted a comprehensive analysis of the chemo-dynamical properties of P180956, leading to the following results:
\begin{enumerate}
    \item The high efficiency and wide spectral coverage of the GHOST instrument (see Figure~\ref{Fig:spectra}) enabled the detection of approximately 20 atomic species (see Figure~\ref{Fig:chems}),
    which is almost twice than what was possible to measure with GRACES \citep{Kielty21,Jeong23,Sestito23,Sestito23Umi,Waller23}. These species provide crucial insights into the origin and chemical properties of P180956 and its formation site.
    \item Theoretical models of supernovae yields suggest that the formation site of P180956 experienced pollution from 2 to 4 low-mass hypernovae ($\lesssim15\msun$) and one intermediate-mass  ($80-120\msun$) fast-rotating ($\sim300\kms$) supernova (see Figure~\ref{Fig:yields}).
    \item These combinations of supernovae events resulted in a composition of $\alpha$-elements with solar Ca and Mg abundances and enhanced Si and Ti abundances (see Figures~\ref{Fig:chems}~and~\ref{Fig:yields}). 
    \item The specific combination of supernovae yields led to low abundances of neutron-capture elements (Sr, Ba, Eu), with a relatively high [Sr/Ba] ratio (see Figure~\ref{Fig:neutron}). This can be explained with the additional s-process channels for Sr production that occur mostly in fast-rotating supernovae.
    \item The kinematical properties of P180956 (see Figure~\ref{Fig:orbit}) suggests the star was likely accreted during the early-assembly phase of the Milky Way. Its [Mg/Mn] ratio is also indicative of its accreted origin (see Figure~\ref{Fig:mgmn}). 
    \item None of the known accreted structures exhibit chemo-dynamical properties  resembling perfectly those of P180956. Only Gaia-Sausage-Enceladus has similar eccentricity, apocentric and pericentric distances to P180956.
    \item P180956 originated in an ancient system chemically similar to  present ultra-faint dwarf galaxies, given the low amount of neutron-capture elements  (see Figures~\ref{Fig:chems}~and~\ref{Fig:neutron}), either accreted alone or dragged in with Gaia-Sausage-Enceladus as its satellite.
\end{enumerate}

The advent of the GHOST high-resolution spectrograph has been invoked by various chemo-dynamical investigations targeting the MW and its satellite systems \citep[\eg][]{Sestito23,Sestito23Umi,Waller23}. This study, along with \citet{Hayes23} and \citet{Dovgal24}, demonstrates that the combination of the Gemini South's large aperture and GHOST's high efficiency and wide spectral coverage is ideal for investigating low-metallicity stars in the Milky Way and nearby systems. The synergy between GHOST and the  Gaia satellite will undoubtedly propel Galactic Archaeological studies forward.

\section*{Acknowledgements}
We acknowledge and respect the l\textschwa\textvbaraccent {k}$^{\rm w}$\textschwa\ng{}\textschwa n peoples on whose traditional territory the University of Victoria stands and the Songhees, Esquimalt and $\ubar{\rm W}$S\'ANE\'C  peoples whose historical relationships with the land continue to this day.

We thank the anonymous referee for their insightful comments, which improved the quality of the manuscript.

FS thanks the Dr. Margaret "Marmie" Perkins Hess postdoctoral fellowship for funding his work at the University of Victoria. KAV thanks the National Sciences and Engineering Research Council of Canada for funding through the Discovery Grants and CREATE programs. AAA acknowledges support from the Herchel Smith Fellowship at the University of Cambridge and a Fitzwilliam College research fellowship supported by the Isaac Newton Trust. NFM gratefully acknowledge support from the French National Research Agency (ANR) funded project ``Pristine'' (ANR-18-CE31-0017) along with funding from the European Research Council (ERC) under the European Unions Horizon 2020 research and innovation programme (grant agreement No. 834148).  TM acknowledges the Spinoza Grant from the Dutch Research Council (NWO) for supporting his research. ES acknowledges funding through VIDI grant "Pushing Galactic Archaeology to its limits" (with project number VI.Vidi.193.093) which is funded by the Dutch Research Council (NWO). This research was supported by the International Space Science Institute (ISSI) in Bern, through ISSI International Team project 540 (The early Milky Way).

This work is based on observations obtained with Gemini South/GHOST, during the commissioning run of September 2022. Based on observations obtained at the international Gemini Observatory, a program of NSF’s NOIRLab, which is managed by the Association of Universities for Research in Astronomy (AURA) under a cooperative agreement with the National Science Foundation. On behalf of the Gemini Observatory partnership: the National Science Foundation (United States), National Research Council (Canada), Agencia Nacional de Investigaci\'{o}n y Desarrollo (Chile), Ministerio de Ciencia, Tecnolog\'{i}a e Innovaci\'{o}n (Argentina), Minist\'{e}rio da Ci\^{e}ncia, Tecnologia, Inova\c{c}\~{o}es e Comunica\c{c}\~{o}es (Brazil), and Korea Astronomy and Space Science Institute (Republic of Korea).

This work has made use of data from the European Space Agency (ESA) mission {\it Gaia} (\url{https://www.cosmos.esa.int/gaia}), processed by the {\it Gaia} Data Processing and Analysis Consortium (DPAC, \url{https://www.cosmos.esa.int/web/gaia/dpac/consortium}). Funding for the DPAC has been provided by national institutions, in particular the institutions participating in the {\it Gaia} Multilateral Agreement.

This research has made use of the SIMBAD database, operated at CDS, Strasbourg, France \citep{Wenger00}. This work made extensive use of \textsc{TOPCAT} \citep{Taylor05}.

\section*{Data Availability}
The data underlying this article are available in the article and in its online supplementary material.

\bibliographystyle{mn2e}
\bibliography{ghost_planar}

\bsp	
\label{lastpage}
\end{document}